\documentstyle[12pt]{article}
\newcommand{\be}{\begin{equation}}
\newcommand{\ee}{\end{equation}}
 
\def\pa{\partial}

\def\a{\alpha}

\def\m{\mu} 
\def\n{\nu}

\def\t{\tau}

\begin{document}
\begin{flushright}
BRX TH 437\\
\end{flushright} 
\begin{center}
{\large Equivalence of Hawking and Unruh Temperatures and Entropies \\ Through
 Flat Space Embeddings\\}

\vspace{0.3cm}
S.Deser and Orit Levin\\
\vspace{0.3cm}

Physics Department, Brandeis University, Waltham, Massachusetts
02454, USA\\
\vspace{1cm}

{\bf Abstract\\}
\end{center}
\begin{quotation}
  We present a unified description of temperature and entropy in spaces with 
either ``true" or
 ``accelerated observer" horizons: In their (higher dimensional) global 
embedding  Minkowski geometries, the relevant detectors have constant
accelerations $a_{G}$; associated with their Rindler horizons are
 temperature $a_{G}/2\pi$ and entropy equal to 1/4 the horizon area.
Both quantities agree with those calculated in the original curved spaces.
 As one example of this equivalence, we
obtain the temperature and entropy of Schwarzschild geometry from its
flat D=6 embedding.\\
\end{quotation}
\vspace{0.5mm}

The relation between Hawking and Unruh effects has been extensively studied,
and the emergence of temperature and entropy due to their respective ``real" or
``accelerated" horizons is well-understood. Recently, it was shown that the
temperatures measured by accelerated detectors in de Sitter (dS) and Anti de
Sitter (AdS) geometries can be obtained \cite{dl} from their corresponding
 constant (Rindler) accelerations in the appropriate global embedding
Minkowski spacetimes (GEMS).  Our purpose here is to
generalize this method to provide a unified kinematical treatment of both
effects in terms of accelerated
motions in the GEMS; any Einstein geometry can be so embedded
\cite{goenner}.  As a first illustration, we obtain the entropy of dS in this
way. We will then derive, as our main example, the
equivalence between temperature and entropy measured by the
usual static detector in Schwarzschild geometry and their values as calculated
from its Rindler-like motion in the (D=6) GEMS.
A more complete discussion of this equivalence, and of its validity for
other important geometries
such as BTZ, Schwarzschild-AdS, Schwarzschild-dS and Riessner-Nordstrom, will be
presented elsewhere. 

Recall first that in flat space, observers with constant acceleration 
of magnitude $a$,
who follow a timelike Killing vector field $\xi$ that encounters
an event horizon, will thereby measure a temperature, $2\pi T = a$.  It is also
well-known that the connection between surface gravity
$k_H$ and temperature,
\begin{equation}
k_H = g_{00}^{1/2} \; 2\pi T
\end{equation}
holds both in black hole spaces and for Rindler motions \cite{wald};
 $x^0$ is the timelike Killing vector of rest detectors
and $k_{H}$ is defined as the horizon value of
$[-\frac{1}{2} (D_\m\xi_\n )^2]^{1/2}$.
In Rindler coordinates, the longitudinal interval is
$ds^2 =L^2 e^{2\zeta} (d\t^2 - d\zeta^2)$ and $\zeta$=const
detectors have acceleration $a=L^{-1}e^{-\zeta}$; they see
an event horizon at $\zeta = -\infty$ ($x^2-t^2=0$), where the Killing
vector $\pa_\t$ is null.  There, $k_H = 1$, hence (1)
immediately reproduces $2\pi T = a$. The entropy of these observers is also
known \cite{lafl}: it is $A/4$, where $A$ is the (transverse) horizon area;
for ``unrestricted" Rindler motion it is in general infinite, but not (as we
shall see) in the GEMS context when real horizons are being represented.

Next, for orientation, we summarize the GEMS 
approach in its original dS/AdS setting. These spaces are hyperboloids
\begin{equation}
\eta_{AB}  z^Az^B = \mp R^2
\end{equation}
in flat, $ds^2 = \eta_{AB} dz^A dz^B$, GEMS.  
Here $A,B = 0 . . D$,  $\eta_{AB} = diag (1,-1..-1, \mp 1)$
and upper/lower signs always refer to dS/AdS respectively.
Now consider
$z^2\! =\!\!...\!=\!z^{D-1}\!=\!0,$ $z^D=Z$ trajectories; these
obey $(z^1)^2 - (z^0)^2 = \pm (R^2-Z^2) \equiv a^{-2}_G$,
and describe constant acceleration Rindler motions of the form 
``$x^2 - t^2 = a^{-2}$ " for $Z$=const.  Such detectors
therefore measure $2\pi T = a_G$, which we rewrite in
terms of the original dS/AdS accelerations $a_D$ as
\begin{equation}
2\pi T = a_G = (\pm R^{-2} + a^2_D)^{1/2} \; .
\end{equation}
This relation is a particular case of the familiar Gauss--Codazzi--Ricci
equation: detectors following a timelike Killing vector $\xi$
in the physical space have acceleration \cite{n}
$a_D = D_{\xi}\xi/ |\xi |^2$ with 
\begin{equation}
a^2_G = a_{D}^2 + \a^2 | \xi |^{-4}
\end{equation}
where $\a$ is the second fundamental form \cite{goenner}; this relation 
extends to cases such as \linebreak Schwarzschild where the GEMS is more 
than one dimension higher.  It should not be inferred from (4) that there
is always a meaningful temperature (in either the original or the embedding 
space), since
$\a^2$ need not always be positive; for example it is  
$\a^2|\xi|^{-4}=-R^{-2}$ for AdS.  After all, the flat space Unruh description
itself is only meaningful when $a^2_G \geq 0$.  We refer
to \cite{dl} for details.
To calculate entropy here, we must restrict the (transverse) horizon area 
integration according
to the definitions of the embedding coordinates. For
 (D=4) dS the three-fold integral over $dz^2dz^3dz^4$ is
restricted by the condition $[(z^2)^2+(z^3)^2+(z^4)^2]^{1/2}=R$, resulting in
the same finite 2-surface value $4\pi R^2$ as calculated directly in D=4.
For the corresponding AdS observers, the coordinate restriction is instead
 $[-(z^2)^2-(z^3)^2+(z^4)^2]^{1/2}=R$ and the area integral is
now infinite. This is not surprising since AdS, like the Rindler wedge, has no
true horizon and its D=4 entropy is also infinite.

Let us now turn to the Schwarzschild metric to exemplify spaces with ``real" 
horizons. Here the
GEMS  \cite{fronsdal}, which covers the usual Kruskal \cite{kruskal}
extension, has D=6 $ds^2=\eta_{AB}dz^{A}dz^{B}$ 
$\eta_{AB}= diag (1,-1,-1,-1,-1,-1)$,  with
\begin{equation}
z^0 = 4m (1-u)^{1/2} \sinh t/ 4m \;\;\;\;
z^1 = 4m (1-u)^{1/2} \cosh t/ 4m 
\end{equation}
\[z^2 = -2m  \int du [u + u^2 +u^3]^{1/2} u^{-2},\;\;
(z^3, z^4, z^5 ) = (x,y,z),\;\;u \equiv 2m/r, \;\; r^2 \equiv
x^2 + y^2 + z^2.\]
This is a global embedding, with extendability
to $r < 2m$. [In incomplete embedding spaces, such as those in \cite{rosen}, 
that cover only the exterior region $r > 2m$, observers see no event
horizon, hence no loss of information or temperature.]  The Hawking 
detectors with constant $\bf r$ in D=4 are here Unruh detectors; their
D=6 motions are the now familiar hyperbolic trajectories
\begin{equation}
(z^1)^2 - (z^0)^2 = 16m^2 (1-u) \equiv a^{-2}_6\; .
\end{equation}
We therefore immediately infer the usual Schwarzschild local Hawking ($T$) and
black hole ($T_0$) temperatures 
\begin{equation}
2\pi T = a_6 = [16m^2(1-u)]^{-1/2}, \;\;\;\;
T_0 = g^{1/2}_{00} T = (8\pi m)^{-1} \; ,
\end{equation} from the purely kinematical GEMS motion's Unruh temperature.

We consider next the equivalence of entropy definitions. Here  we must 
restrict the four-fold transverse integration
over $dz^2...dz^5$ to the underlying manifold using the embedding definition
(5), including the condition $r=2m$ implied by $(z^0)^2-(z^1)^2=0$, to be
enforced by introduction of the obvious delta function with $r$ expressed as 
$[(z^3)^2+(z^4)^2+(z^5)^2]^{1/2}$. This just restricts the $z^3, z^4, z^5$
volume to the surface of $r=2m$, while the $z^2$ integral is just equal to 
1; then it is a simple matter to verify that
the integration gives the transverse area $4\pi (2m)^2$, precisely that of the
Schwarzschild horizon, and thereby establish the equivalence.

We have here exploited the GEMS approach to unify the definitions of
temperatures and entropies whatever the origin of their event horizons.
The ``equivalence" that emerged is closely related to its usual Einstein
 meaning, even though we are dealing here with nonlocal properties.
From the definitions of temperature and
entropy with flat space accelerated motions, the GEMS
mechanism ensures the correspondence with curved space horizons as well, at
least for sufficiently simple horizon structures where detectors are mapped
into Rindler ones; the quantitative equivalence
is then essentially guaranteed.
We hope to return to other, ``non-horizon", uses of GEMS,
for example those involving rotation and superradiance.\\  

This work was supported by NSF grant PHY-9315811;
 OL thanks the Fischbach Foundation for a fellowship.

\end{document}